\documentclass[conference]{IEEEtran}

\usepackage{times}
\usepackage{epsfig}
\usepackage[noadjust]{cite}
\usepackage{amsmath}
\usepackage{graphicx}
\usepackage{listings,algorithm}
\usepackage{multirow}
\usepackage{array}
\usepackage{url}
\usepackage{color}
\usepackage[normalem]{ulem}
\usepackage{amssymb}
\usepackage[justification=centering,font=footnotesize]{caption}
\usepackage{flushend}
\usepackage{algpseudocode}

\IEEEoverridecommandlockouts

\begin{document}

\newcommand{\LO}{$\mathtt{LO}$}
\newcommand{\HI}{$\mathtt{HI}$}

\algrenewcomment[1]{\(\triangleright\) #1}

\title{\Large\bf Mixed-Criticality Scheduling with I/O}
\author{Eric Missimer, Katherine Zhao and Richard West\\[0.1in]
  \em  Computer Science Department \\
  Boston University \\
  Boston, MA 02215\\
  Email: \{missimer,kzhao,richwest\}@cs.bu.edu \\[0.1in]
}

\date{}

\maketitle
\thispagestyle{plain}
\pagestyle{plain}

\begin{abstract}
This paper addresses the problem of scheduling tasks with different
criticality levels in the presence of I/O requests. In mixed-criticality
scheduling, higher criticality tasks are given precedence over those of lower
criticality when it is impossible to guarantee the schedulability of all
tasks. While mixed-criticality scheduling has gained attention in recent
years, most approaches typically assume a periodic task model.  This
assumption does not always hold in practice, especially for real-time and
embedded systems that perform I/O.  In prior work, we developed a
scheduling technique in the Quest real-time operating system, which integrates
the time-budgeted management of I/O operations with Sporadic Server scheduling
of tasks. This paper extends our previous scheduling approach with support for
mixed-criticality tasks and I/O requests on the same processing core.  Results
show that in a real implementation the mixed-criticality scheduling method
introduced in this paper outperforms a scheduling approach consisting of only
Sporadic Servers.

\end{abstract}

\section{Introduction}
\label{sect:intro}
Mixed-criticality scheduling orders the execution of tasks of different
criticality levels. Criticality levels are based on the consequences of a task
violating its timing requirements, or failing to function as specified. For
example, DO-178B is a software certification used in avionics, which specifies
several assurance levels in the face of software failures. These assurance
levels range from catastrophic (e.g., could cause a plane crash) to
non-critical when they have little or no impact on aircraft safety or overall
operation.
Mixed-criticality scheduling was first introduced by Vestal
(2007)~\cite{Vestal07}.  Later, Baruah, Burns and Davis
(2011)~\cite{BaruahBuDa11} introduced Adaptive Mixed-Criticality (AMC)
scheduling.  The work presented in this paper builds upon AMC to extend it for
use in systems where tasks make I/O requests. This is the first paper to address
the issue of I/O scheduling in an Adaptive Mixed-Criticality scenario.  Our
approach to AMC with I/O is based on experience with our in-house real-time
operating system, called Quest~\cite{quest-webpage}.

\begin{figure}[!ht]
  \vspace{-0.1in}
  \centering
  \includegraphics[width=0.44\textwidth]{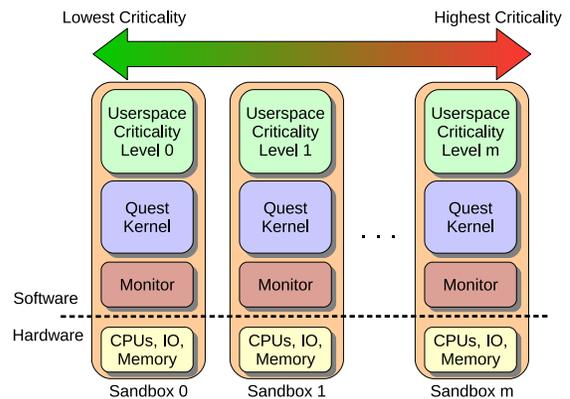}
  \vspace{0.1in}
  \caption{Mixed-Criticality Levels Across Separate Quest-V Sandboxes}
  \label{fig:quest-v}
  \vspace{-0.08in}
\end{figure}

\begin{figure}[!ht]
  \vspace{-0.1in}
  \centering
  \includegraphics[width=0.3\textwidth]{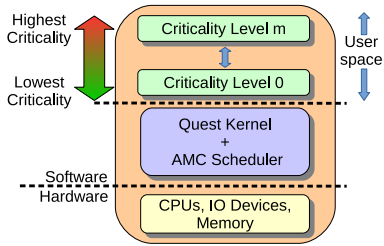}
  \vspace{0.1in}
  \caption{Quest Support for Mixed-Criticality Scheduling}
  \label{fig:quest-v-amc}
\end{figure}

Quest can be configured to have two privilege levels, with the more privileged
kernel being separated from a less privileged user-space. This is similar to
traditional operating systems such as Linux. In contrast, an alternative
system configuration, called Quest-V, supports three privilege levels. The
third privilege level in Quest-V is more trusted than the kernel, and operates
as a lightweight virtual machine monitor, or hypervisor. Unlike with
traditional virtual machine systems, Quest-V uses its most trusted privilege
level to {\em partition} resources amongst (guest) sandbox domains. Each
sandbox domain then manages its own resources independently and in isolation
of other sandbox domains, without recourse to a hypervisor. This leads to a
far more efficient design, where virtualization overheads are almost entirely
eliminated. It has been shown in prior work that it is possible to dedicate
separate tasks of different criticality levels to different sandboxes in
Quest-V~\cite{LiWeMi2014}.  This is demonstrated in Figure~\ref{fig:quest-v}.
Note that each sandbox has a different criticality level with level 0 being
the least critical.  However, Quest-V has thus far not considered tasks of
different criticality levels {\em within} the same sandbox and, hence, for
scheduling on the same (shared) processor cores. In this paper, we show how to
integrate the processing of I/O events and tasks in an Adaptive
Mixed-Criticality~\cite{BaruahBuDa11} scheduling framework built within the
Quest kernel. We extend Quest with support for {\em mode changes} between
different criticality levels. This enables components of different critically
levels to coexist in a single Quest-V sandbox or in a traditional Quest
system, as depicted in Figure~\ref{fig:quest-v-amc}.

Previous mixed-criticality analysis assumes that all jobs in the system are
scheduled under the same policy, typically as periodic tasks.  However, as
previously shown by Danish, Li and West~\cite{DanishLiWe11}, using the same
scheduling policy for both task threads and bottom half interrupt
handlers\footnote{We use the Linux terminology, where the top half is the
non-deferrable work that runs in interrupt context, and the bottom half is the
deferrable work executed in a thread context after the top half.} results in
lower I/O performance and larger overheads.  Specifically, the authors
compared the Sporadic Server (SS)~\cite{Sprunt90} model for both main threads
and bottom half interrupt handlers to using Sporadic Servers for main threads
and Priority Inheritance Bandwidth-preserving Servers (PIBS) for bottom half
threads.  The results showed that by using PIBS for interrupt bottom half
threads, the scheduling overheads are reduced and I/O performance is
increased.  The details of PIBS will be discussed in
Section~\ref{sect:ss_and_pibs}.

The contributions of this paper include a mixed-criticality analysis assuming
threads are scheduled using either the Sporadic Server or PIBS scheduling
model. It is shown that while a system of Sporadic Servers and PIBS has a
slightly lower schedulability than a system of only Sporadic Servers from a
theoretical point of view, in practice a real implementation of both scheduling
policies results in Sporadic Server and PIBS outperforming a system of only
Sporadic Servers.

The rest of the paper is organized as follows.  Section~\ref{sect:ss_and_pibs}
provides the necessary background information on Sporadic Servers and PIBS and
introduces a response time analysis for them.  Next, Section~\ref{sect:amc}
briefly discusses the Adaptive Mixed-Criticality (AMC) model.
Section~\ref{sect:amc_ss_and_pibs} contains the AMC scheduling analysis for a
system of Sporadic and Priority Inheritance Bandwidth Preserving Servers.
Section~\ref{sect:implementation} discusses the implementation details of our
AMC approach in the Quest operating system.  Section~\ref{sect:experimental}
discusses experimental results, while related work is described in
Section~\ref{sect:related}.  Finally, conclusions are discussed in
Section~\ref{sect:conclusions}.

\section{Sporadic Server and PIBS}
\label{sect:ss_and_pibs}
Sporadic Servers (SS)~\cite{Sprunt90} and Priority Inheritance
Bandwidth-preserving Servers (PIBS)~\cite{DanishLiWe11} are the two scheduling
models used in the Quest real-time operating system~\cite{quest-webpage}.
Sporadic Servers are specified using a budget capacity, $C$, and period
$T$. By default, the Sporadic Server with the smallest period is given highest
priority, which follows the rate-monotonic policy~\cite{LiuLa73}. The main
tasks in Quest run on Sporadic Servers, thereby guaranteeing them a minimum
share of CPU time every real-time period.  Replenishment lists are used to
track the consumption of CPU time and when it is eligible to be re-applied to
the corresponding server.

PIBS uses a much simpler scheduling method which is more appropriate for the
short execution times associated with interrupt bottom half threads.  A PIBS
is specified by a utilization, $U$.  A PIBS always runs on behalf of a
Sporadic Server and inherits both the priority and period of the Sporadic
Server.  For example, the PIBS running in response to a device interrupt would
run on behalf of the Sporadic Server that requested the I/O action to be
performed.  The capacity of a PIBS is calculated as $C{=}U{\times}T$, where
$T$ is the period of the Sporadic Server.

\begin{figure}[h]
  \centering
  \includegraphics[width=0.5\textwidth]{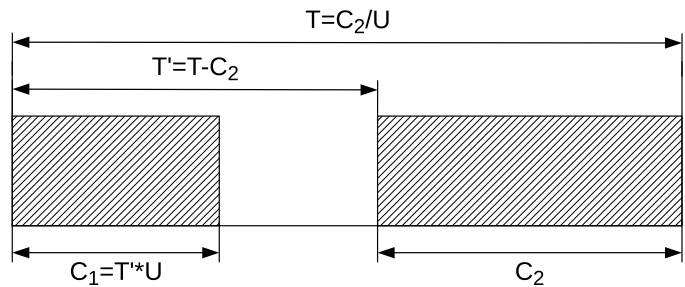}
  \caption{PIBS Server Utilization}
  \label{fig:PIBS}
\end{figure}

As with Sporadic Servers, PIBS uses replenishments but instead of a list there
is only a {\em single} replenishment.  When a PIBS has executed
$C_\mathtt{actual}$, its next replenishment is set to $t+C_\mathtt{actual}/U$,
where $t$ is the time the PIBS started its most recent execution.  A PIBS cannot
execute again until the next replenishment time regardless of whether it has
utilized its entire budget or not. Since a PIBS uses only one replenishment
value rather than a list, it is beneficial for scheduling short-lived interrupt
service routines that would otherwise fragment a Sporadic Server's budget into
many small replenishments. The replenishment method of a PIBS limits its maximum
utilization within any sliding window of size $T$ to $\left(2-U\right)U$.  This
occurs when the PIBS first runs for $C_{1}{=}U(T-UT)$ and then again for
$C_{2}{=}UT$.  This is demonstrated in Figure~\ref{fig:PIBS}:

\begin{align*}
\frac{C_{\mathtt{1}} + C_{\mathtt{2}}}{T}
&=\frac{\left(T^\prime*U\right)+C_{\mathtt{2}}}{T} \\
&=\frac{\left(T-C_{\mathtt{2}}\right)*U+C_{\mathtt{2}}}{T} \\
&=\frac{\left(C_{\mathtt{2}}/U-C_{\mathtt{2}}\right)*U+C_{\mathtt{2}}}{C_{\mathtt{2}}/U} \\
&=\left(2-U\right)U
\normalsize
\end{align*}

\begin{figure}[h]
  \vspace{0.1in}
  \centering
  \includegraphics[width=0.44\textwidth]{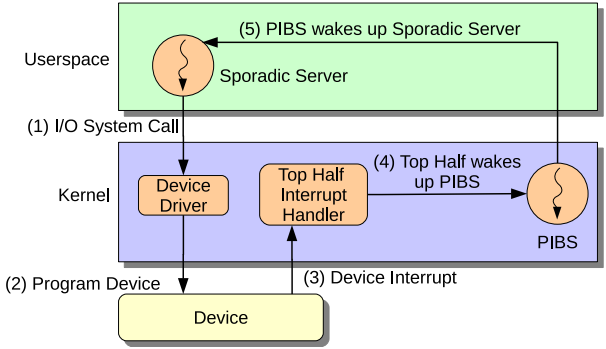}
  \vspace{0.1in}
  \caption{Sporadic Server and PIBS Interaction}
  \label{fig:vcpus}
  \vspace{-0.3in}
\end{figure}

The interaction between Sporadic Servers and PIBS is depicted in
Figure~\ref{fig:vcpus}.  First, the Sporadic Server initiates an I/O related
system call (Step 1).  The system call invokes the associated device driver,
which programs the device (Step 2).  The device will eventually initiate an
interrupt which is handled by the top half interrupt handler (Step 3).  The
top half interrupt handler will acknowledge the interrupt and wake up one of
the PIBS to handle the majority of the work associated with the interrupt
(Step 4).  Note that although the figure shows PIBS run at kernel-level, they
could just as well be associated with user-space threads if the system granted
such privileges.  Finally, after a PIBS finishes executing it will wake up the
corresponding Sporadic Server, assuming the Sporadic Server was blocked on an
I/O request (Step 5).  A more detailed description of the scheduling of PIBS
and Sporadic Servers can be found in prior work (2011)~\cite{DanishLiWe11}.

The advantages of using PIBS for bottom half interrupt handling include lower
scheduling overhead and no delayed execution due to replenishment list
fragmentation.  The short execution time of bottom half interrupt handlers can
cause a Sporadic Server to complete execution before exhausting its available
capacity. This leads to a fragmented replenishment list. In practice, the lists
are finite in length because of memory constraints and to limit scheduling
overhead. When a replenishment list is full, items are merged to make space for
new replenishments. This causes the available budget to be
deferred~\cite{StanovichBaWa10}, and the effective utilization of the Sporadic
Server can drop below its specified value. This in turn results in deadlines
being missed.  In contrast, PIBS have only a single replenishment list item and
a different policy for how the replenishment is posted, which prevents a drop in
their effective utilization and lower scheduling overhead.

\begin{figure}[h]
  \centering
  \includegraphics[width=0.5\textwidth]{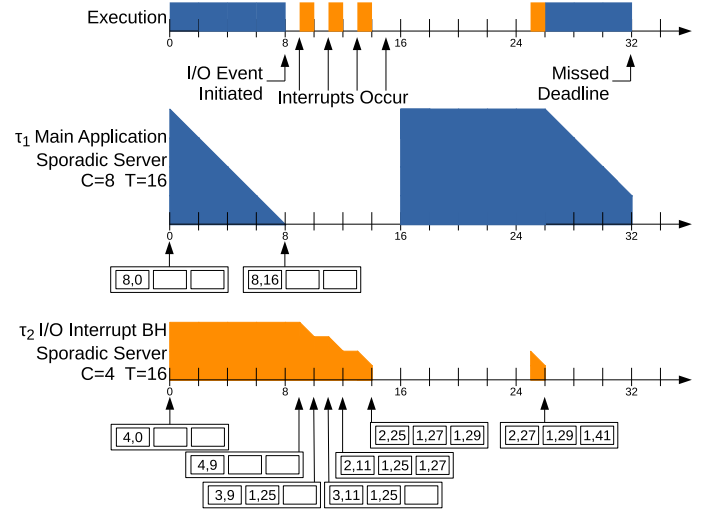}
  \vspace{0.05in}
  \caption{Example Task and I/O Scheduling using Sporadic Servers}

  \label{fig:gantt_ss_only}
\end{figure}

Figure~\ref{fig:gantt_ss_only} shows an example of replenishment list
fragmentation.  The first task, $\tau_1$, begins execution at time $t{=}0$ and
continues execution for eight time units.  $\tau_1$ utilized its entire
capacity at $t{=}8$ so a single single replenishment item is posted for 8 time
units of capacity at time $t{=}16$.  The replenishment is posted at $t{=}16$
because $\tau_1$ started execution at time $t{=}0$ and has a period of 16.
Right at the completion of its execution $\tau_1$ initiates an I/O related
event, e.g. a read.  Suppose this I/O event causes four interrupts to
occur.  Each interrupt initiates a bottom half thread that takes one time unit
of computation to complete.  $\tau_1$ will require all four bottom half
interrupt handlers to complete execution before it can continue execution,
e.g. it is blocking on the read.  $\tau_2$ is the task responsible for
handling these bottom half interrupt handlers.  The first interrupt occurs at
time $t{=}9$ and is immediately handled by $\tau_2$.  Note that at time
$t{=}9$ the time of the head replenishment list item is updated from zero to
nine.  This is to ensure that when a replenishment item is posted after the
task blocks or depletes its budget the replenishment item is posted at the
correct time.  Once $\tau_2$ completes execution of the bottom half interrupt
handler it blocks as it waits for another I/O interrupt to occur.

When $\tau_2$ blocks it posts a replenishment item for the capacity that it
used.  Since it used 1 time unit of capacity and started executing at time
$t{=}9$ a replenishment item of 1 time unit is posted at time $t{=}25$.  At
time $t=11$ another interrupt occurs, waking up $\tau_2$ for another time
unit.  The time of the first replenishment list item is updated to 11 to
reflect that the Sporadic Server started execution at time $t{=}11$.  After
handling the bottom half interrupt handler, another replenishment item for one
time unit is posted, this time at time $t{=}27$.  When the third interrupt
occurs $\tau_2$ again executes for 1 time unit.  However, when $\tau_2$
attempts to post a replenishment item for the one unit of capacity used it
cannot since its replenishment list is full.\footnote{For the sake of this
example the replenishment list size is three.  In practice, a larger
replenishment list size would be chosen but, regardless, fragmentation and
capacity postponement can occur~\cite{DanishLiWe11}.}  In order to ensure that
$\tau_2$ does not adversely affect other running tasks, its remaining capacity
of one time unit is merged with the next replenishment list item, which in
this example is at $t{=}25$.  This results in the available capacity for
$\tau_2$ being zero, leaving it unable to immediately handle the interrupt
that occurs at time $t{=}15$.  Instead, the execution of the interrupt is
delayed and completes only at time $t{=}26$.  Meanwhile, $\tau_1$, which had
the capacity to execute at time $t{=}16$, is blocking waiting for completion of
the fourth interrupt handler.  $\tau_1$ begins execution at time $t{=}26$ but
that leaves only six time units until the deadline at time $t{=}32$, instead
of the eight required to complete execution.

\begin{figure}[h]
  \vspace{-0.15in}
  \centering
  \includegraphics[width=0.5\textwidth]{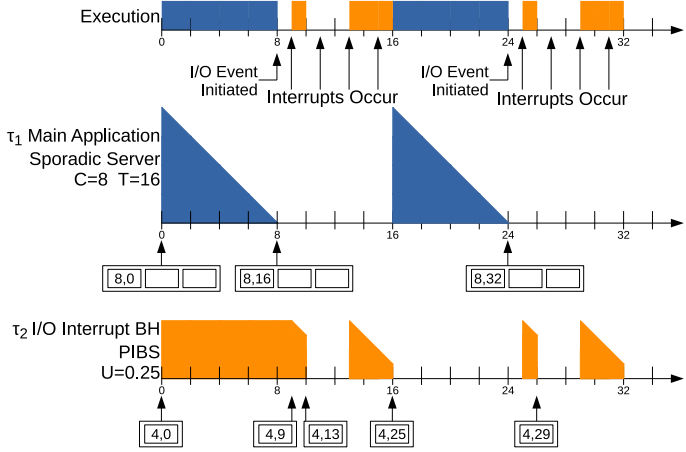}
  \vspace{0.05in}
  \caption{Example Task and I/O Scheduling using Sporadic Servers \& PIBS}

  \label{fig:gantt_ss_pibs}
  \vspace{-0.15in}
\end{figure}

Figure~\ref{fig:gantt_ss_pibs} shows a similar scheduling scenario. However,
this time the interrupt bottom halves are handled by a PIBS.  As with the
previous scenario, $\tau_1$ initiates an I/O related event at time $t{=}8$ and
blocks until the completion of the event.  The first interrupt occurs at time
$t{=}9$ and is immediately handled by PIBS.  As with the Sporadic Server, the
time in the replenishment list item is updated to reflect when the PIBS
started execution.  Once the event is handled, the PIBS posts a single
replenishment item at time $t{=}13$.  This is because $\tau_2$ is running on
behalf of $\tau_1$ so it inherits both the priority and period of $\tau_1$.
Since $\tau_2$ executed for only 25\% of its available four time units of
capacity the replenishment is posted 25\% of its period from when it started
execution.  The second interrupt occurs at time $t{=}11$ but its execution is
deferred until $\tau_2$ has available capacity.  At time $t{=}13$ the third
interrupt arrives and $\tau_2$ has the capacity to handle both it and the
previous interrupt.  Finally, the fourth interrupt arrives at time $t{=}15$,
which can also be handled by $\tau_2$.  Since $\tau_2$ has executed for 75\%
of its available capacity a replenishment is posted twelve time units after it
started execution, at $t{=}25$.  This permits $\tau_1$ to continue execution
at time $t{=}16$.  The pattern then repeats itself.  This simple example
demonstrates the advantages of PIBS for bottom half threads compared to
Sporadic Servers.  Finally, note that even if the replenishment list in the
first example had been long enough to avoid the delayed budget, the Sporadic
Server would have experienced twice as much context switching overhead
compared to the equivalent PIBS.

Note that in the first example, if a different policy for handling a full
Sporadic Server replenishment list had been chosen, $\tau_2$ might have
completed in time for $\tau_1$ to finish before its deadline.  For example, if
the later replenishment items were merged instead of the head replenishment
item, $\tau_2$ would have had one remaining time unit of capacity to handle
the last bottom half interrupt handler.  However, as more interrupts occur,
this temporary fix will no longer work as more capacity is delayed further in
time.

\subsection{Response Time Analysis for SS and PIBS}

In order to perform an Adaptive Mixed-Criticality analysis for a combined
Sporadic Server and PIBS system, the response time analysis equation of the
system must be derived.  First, under the assumption that a Sporadic Server
can be treated as an equivalent periodic task~\cite{Sprunt90}, the response
time equation for task $\tau_i$ in a system of only Sporadic Servers is the
following:
\[
R_i = C_i +
\sum_{\tau_j\in\mathbf{hp}\left(i\right)}{\left\lceil\frac{R_i}{T_j}\right\rceil
  C_j}
\]
where $\mathbf{hp}\left(i\right)$ is the set of tasks of equal or higher
priority than task $\tau_i$. Second, due to the worst-case phasing of a
combined system of PIBS and Sporadic Servers, a PIBS utilization bound of
$\left(2-U\right)U$ cannot repeatedly occur.  The worst case phasing can
result in at most an additional capacity (i.e., execution time) of $\left(T_q{-}
T_q U_k\right)U_k$ for PIBS $\tau_k$ assigned to the Sporadic Server
$\tau_q$. This is only possible if PIBS blocks waiting on I/O before consuming
its full budget capacity.
Therefore, a
tighter upper-bound on the interference a PIBS can cause is:
\begin{align*}
I^q_k\left(t\right)
&= \left(T_q{-}T_q U_k\right)U_k + \left\lceil \frac{t}{T_q} \right\rceil T_q U_k \\
&= \left(1-U_k\right)T_q U_k + \left\lceil \frac{t}{T_q} \right\rceil T_q U_k \\
&= \left(1 + \left\lceil \frac{t}{T_q} \right\rceil - U_k \right) T_q U_k
\end{align*}
This can be incorporated into the response time analysis of Sporadic Server
$\tau_i$, in a system consisting of both Sporadic Servers and PIBS, in the
following way:
\begin{align} \label{eq:ss_rtb}
R_i = C_i &+ \sum_{\tau_j\in\mathbf{hp}\left(i\right)} \left\{ \left\lceil
    \frac{R_i}{T_j} \right\rceil C_j\right\} \notag \\ & +
\sum_{\tau_k \in \mathbf{ps}}\max_{\tau_q \in
  \mathbf{hip}\left(i\right)}\left\{I_k^q\left(R_i\right)\right\}
\end{align}
Where $\mathbf{ps}$ is the set of all PIBS and \mbox{$\mathbf{hip}\left(i\right)
  {=} \mathbf{hp}\left(i\right)\cup\left\{\tau_i\right\}$}, i.e. the set
containing $\tau_i$ and all tasks with equal or higher priority than task
$\tau_i$.  This is necessary as the PIBS can be running on behalf of task
$\tau_i$.  In general, there is no a-priori knowledge about which PIBS runs for
which Sporadic Server.  Therefore, the Sporadic Server, $\tau_q$ that maximizes
the interference caused by the PIBS must be considered.  If such a-priori knowledge
existed, it could be used to reduce the possible set of Sporadic Servers on
  behalf of which a PIBS could be executing. However, without such knowledge all possible
Sporadic Server tasks of equal or higher priority must be considered.

The response time analysis for a PIBS is therefore dependent on the associated
Sporadic Server.  The response time analysis for PIBS $\tau_p$ when assigned to
Sporadic Server $\tau_s$ is:
\begin{align} \label{eq:pibs_rta}
  _{s}R_p = &\left(2-U_p\right)U_pT_s +
  \sum_{\tau_j\in\mathbf{hip}\left(s\right)} \left\{ \left\lceil \frac{_{s}R_p}{T_j}
    \right\rceil C_j\right\} \notag \\ &+
  \sum_{\substack{\tau_k \in \mathbf{ps} \setminus \left\{\tau_p\right\}}}
  \max_{\tau_q \in
    \mathbf{hip}\left(s\right)}\left\{I_k^q\left(_{s}R_p\right)\right\}
\end{align}
Note that $\left(2{-}U_p\right)U_pT_s$ is the maximum execution time of the PIBS
over a time window of $T_s$, i.e.  $I_p^s\left(T_s\right) {=}
\left(2{-}U_p\right)U_pT_s$.  Besides the first terms differing,
Equation~\ref{eq:pibs_rta} differs from Equation~\ref{eq:ss_rtb} in that
$\mathbf{hip}\left(s\right)$ is used instead of $\mathbf{hp}\left(s\right)$
for the set of Sporadic Servers.  This is because Sporadic Server $\tau_s$
must be included as it has an equal priority to PIBS $\tau_p$ when $\tau_p$ is
running on behalf of $\tau_s$.  Also, the summation over all PIBS does not
include PIBS $\tau_p$ when determining its response time.  If
$_{s}R_p{\leq}T_s$, for each and every Sporadic Server $\tau_s$ that $\tau_p$
can be assigned to, then $\tau_p$ will never miss a deadline.

\section{Background: AMC Scheduling}
\label{sect:amc}
This section will provide the necessary background information on Adaptive
Mixed-Criticality (AMC) scheduling~\cite{BaruahBuDa11} to understand the
analysis in Section~\ref{sect:amc_ss_and_pibs}.  A more detailed analysis can
be found in Baruah (2011)~\cite{BaruahBuDa11}.

In AMC, a task $\tau_i$ is defined by its period, deadline, a vector of
computation times and a criticality level,
$\left(T_i, D_i, \vec{C}_i, L_i\right)$.  In the simplest case,
$L_i {\in} \left\{\mathtt{LO},\mathtt{HI}\right\}$, i.e. there are two
criticality levels \LO{} and \HI{} where $\mathtt{HI}{>}\mathtt{LO}$.  For tasks
for which $L{=}\mathtt{LO}$, $C\left(HI\right)$ is not defined as there are no
\HI{}-criticality versions of these tasks to execute.  For \HI{}-criticality
tasks $C\left(\mathtt{HI}\right){\geq} C\left(\mathtt{LO}\right)$.  The system
also has a criticality level and it initially starts in the \LO{}-criticality
mode.  While running in the \LO{}-criticality mode, both \LO{}-
and \HI{}-criticality tasks execute, and while running in \HI{}-criticality
mode, only
\HI{}-criticality tasks execute.  If a high criticality task exhausts its
$C\left(LO\right)$ before finishing its current job, the system switches into
the \HI{}-criticality mode and suspends all \LO{}-criticality tasks.  This
requires a signaling mechanism available to tasks to signal that they have
completed execution of a specific job instance.

The schedulability test for AMC consists of three parts: 1) the schedulability
of the tasks when the system is in the \LO{}-criticality state, 2) the
schedulability of the tasks when the system is in the \HI{}-criticality state
and 3) the schedulability of the tasks during the mode change from
\LO{}-criticality to \HI{}-criticality.  The first two are simple and can be
handled with the traditional response time analysis, taking into account the
appropriate set of tasks and worst case execution times.  Specifically, the
response time analysis for each task $\tau_i$ when the system is in
the \LO{}-criticality state is:
\[
R_i^{\mathtt{LO}} = C_i\left(\mathtt{LO}\right) + \sum_{\tau_j\in
\mathbf{hp}\left(i\right)}{\left\lceil\frac{R_i^{\mathtt{LO}}}{T_j}\right\rceil
C_j\left(\mathtt{LO}\right)}
\]
and the response time analysis for the \HI{}-criticality state is:
\[
R_i^{\mathtt{HI}} = C_i\left(\mathtt{HI}\right) + \sum_{\tau_j\in
\mathbf{hpH}\left(i\right)}{\left\lceil\frac{R_i^{\mathtt{HI}}}{T_j}\right\rceil
C_j\left(\mathtt{HI}\right)}
\]
where $\mathbf{hpH}\left(i\right)$ is the set of all high-criticality tasks with
a priority higher than or equal to that of task $\tau_i$.

What remains is whether all \HI{}-criticality tasks will meet their deadlines
during the mode change from \LO{}-criticality to \HI{}-criticality.  Baruah,
Burns and Davis provided two sufficient but not complete scheduling tests for
the criticality mode, i.e. the tests will not admit task sets that are not
schedulable but may reject task sets that are schedulable.  The first is
AMC-rtb (response time bound) which derives a new response time analysis
equation for the mode change.  The second is AMC-max which derives an
expression for the maximum interference a \HI{}-criticality task can
experience during the mode change.  AMC-max iterates over all possible points
in time where the interference could increase, taking the maximum of these
points.  AMC-max is more computationally expensive than AMC-rtb but dominates
AMC-rtb by permitting certain task sets that AMC-rtb rejects, and accepting
any task set that AMC-rtb accepts.  Both tests use Audsley's
priority-assignment algorithm~\cite{Audsley01}, as priorities that are
inversely related to period are not optimal for AMC~\cite{Vestal07,
BaruahBuDa11}.

In this paper, we focus on the use of AMC-rtb for response time analysis of a
system with Sporadic Servers and PIBS. This is because of the added expense
incurred by AMC-max, which must iterate over all time points
when \LO{}-criticality tasks are released. 

The AMC-rtb analysis starts with a modified form of the traditional periodic
response time analysis:

\begin{align} \label{eq:amc_rtb}
R_i^* = C_i + \sum_{\tau_j \in
\mathbf{hp}\left(i\right)}{\left\lceil\frac{R_i^*}{T_j}\right\rceil
C_j\left(\min\left(L_i,L_j\right)\right)}
\end{align}

Where $\min\left(L_i,L_j\right)$ returns the lowest criticality level passed to
it, e.g. in the case of a dual-criticality level system, \HI{} is only returned
if both arguments are \HI{}.  The use of $\min$ implies that we only consider
criticality levels equal to or less than the criticality level of $\tau_i$.  If
we divide the higher priority tasks by criticality level, we obtain
the following:
\begin{align}
\label{equ:amc_der_2}
R_i^* = C_i&+ \sum_{\tau_j \in
\mathbf{hpH}\left(i\right)}{\left\lceil\frac{R_i^*}{T_j}\right\rceil
C_j\left(\min\left(L_i,L_j\right)\right)} \notag \\ &+ \sum_{\tau_j \in
\mathbf{hpL}\left(i\right)}{\left\lceil\frac{R_i^*}{T_j}\right\rceil
C_j\left(\mathtt{LO}\right)}
\end{align}
Where $\mathbf{hpL}\left(i\right)$ is the set of all \LO{}-criticality tasks
with a priority higher than or equal to the priority of task $\tau_i$.  The
$\min$ in the third term is replaced with \LO{} as we know $L_j{=}\mathtt{LO}$.
Since we are only concerned with high priority tasks after the mode change,
i.e. $L_i{=}\mathtt{HI}$, Equation~\ref{equ:amc_der_2} becomes:
\begin{align}
R_i^* = C_i\left(\mathtt{HI}\right)&+ \sum_{\tau_j \in
\mathbf{hpH}\left(i\right)}{\left\lceil\frac{R_i^*}{T_j}\right\rceil
C_j\left(\mathtt{HI}\right)} \notag \\ & + \sum_{\tau_j \in
\mathbf{hpL}\left(i\right)}{\left\lceil\frac{R_i^*}{T_j}\right\rceil
C_j\left(\mathtt{LO}\right)}
\end{align}
Finally, the response time bound can be tightened even further by recognizing
that \LO{}-criticality tasks can only interfere with \HI{}-criticality tasks
before the change has occurred.  With this observation the final AMC response
time bound equation is:
\begin{align}
R_i^* = C_i\left(\mathtt{HI}\right)&+ \sum_{\tau_j \in
\mathbf{hpH}\left(i\right)}{\left\lceil\frac{R_i^*}{T_j}\right\rceil
C_j\left(\mathtt{HI}\right)} \notag \\ &+ \sum_{\tau_j \in
\mathbf{hpL}\left(i\right)}{\left\lceil\frac{R_i^\mathtt{LO}}{T_j}\right\rceil
C_j\left(\mathtt{LO}\right)}
\end{align}

\subsection{\LO{}-criticality tasks running in the \HI{}-criticality mode}

Burns and Baruah~\cite{BurnsBa13} provide an extension to AMC that permits
lower criticality tasks to continue execution in the \HI{}-criticality state.
This extension is used in our AMC model with support for I/O, which is briefly
summarized as follows:

If \LO{}-criticality tasks are allowed to continue execution in the
\HI{}-criticality mode at a lower capacity, the following is the response time
for a \HI{}-criticality task $\tau_i$:
\begin{align} \label{eq:lower_tasks_hi}
  R_i^* = C_i &+ \sum_{\tau_j \in
    \mathbf{hpH}\left(i\right)}{\left\lceil\frac{R_i^*}{T_j}\right\rceil
    C_j\left(\mathtt{HI}\right)} \notag \\ &+ \sum_{\tau_j \in
    \mathbf{hpL}\left(i\right)}{\left\lceil\frac{R_i^\mathtt{LO}}{T_j}\right\rceil
    C_j\left(\mathtt{LO}\right)} \notag \\ & + \sum_{\tau_j \in
    \mathbf{hpL}\left(i\right)}{\left(\left\lceil\frac{R_i^*}{T_j}\right\rceil -
      \left\lceil\frac{R_i^\mathtt{LO}}{T_j}\right\rceil\right)
    C_j\left(\mathtt{HI}\right)}
\end{align}
The final term in Equation~\ref{eq:lower_tasks_hi} expresses the maximum
number of times the \LO{}-criticality task will be released multiplied by its
{\em smaller}\footnote{For \LO{}-criticality tasks that can execute
in \HI{}-criticality mode, $C\left(\mathtt{LO}\right){>}
C\left(\mathtt{HI}\right)$, whereas for \HI{}-criticality tasks  $C\left(\mathtt{HI}\right){\geq}
C\left(\mathtt{LO}\right)$.}
\HI{}-criticality execution time.  While Equation~\ref{eq:lower_tasks_hi} also
applies to \LO{}-criticality tasks that continue running after the mode
change, a tighter bound is possible.  Specifically, if a \LO{}-criticality
task has already run for $C\left(\mathtt{HI}\right)$ before the mode change
then it has met its \HI{}-criticality requirement.  Therefore,
$R_i^\mathtt{LO}$ can be replaced with a smaller value for \LO{}-criticality
tasks.  To this end $R_i^{\mathtt{LO}*}$ is defined as the following:
\begin{align}
R_i^{\mathtt{LO}*} = \min\left(C_i\left(\mathtt{LO}\right), C_i\left(\mathtt{HI}\right)\right) + \notag \\ \sum_{\tau_j\in
\mathbf{hp}\left(i\right)}{\left\lceil\frac{R_i^{\mathtt{LO*}}}{T_j}\right\rceil
C_j\left(\mathtt{LO}\right)}
\end{align}
Note that $R_i^{\mathtt{LO}*}{=}R_i^{\mathtt{LO}}$ if $L_i{=}\mathtt{HI}$ and
$R_i^{\mathtt{LO}*}{\leq}R_i^{\mathtt{LO}}$ if $L_i{=}\mathtt{LO}$, as
\LO{}-criticality tasks will have a smaller capacity in the \HI{}-criticality
mode.  Therefore, Equation~\ref{eq:lower_tasks_hi} can be replaced with the
following more general equation that is tighter for \LO{}-criticality tasks:
\begin{align} \label{eq:lower_tasks_all}
  R_i^* = &C_i+ \sum_{\tau_j \in
    \mathbf{hpH}\left(i\right)}{\left\lceil\frac{R_i^*}{T_j}\right\rceil
    C_j\left(\mathtt{HI}\right)} \notag \\ &+ \sum_{\tau_j \in
    \mathbf{hpL}\left(i\right)}{\left\lceil\frac{R_i^{\mathtt{LO}*}}{T_j}\right\rceil
    C_j\left(\mathtt{LO}\right)} \notag \\ & + \sum_{\tau_j \in
    \mathbf{hpL}\left(i\right)}{\left(\left\lceil\frac{R_i^*}{T_j}\right\rceil -
      \left\lceil\frac{R_i^{\mathtt{LO}*}}{T_j}\right\rceil\right)
    C_j\left(\mathtt{HI}\right)}
\end{align}

In Section~\ref{sect:amc_ss_and_pibs} we will use both AMC models described in
this section to derive an AMC model for a system that includes Priority
Inheritance Bandwidth-Preserving Servers.

\section{AMC Sporadic Server and PIBS Scheduling}
\label{sect:amc_ss_and_pibs}
This section describes the system model for I/O Adaptive Mixed-Criticality
(IO-AMC), comprising both Sporadic Servers and Priority Inheritance
Bandwidth-Preserving Servers (PIBS).  IO-AMC focuses on the scheduling of I/O
events and application threads in a mixed-criticality setting. Based on the
IO-AMC model, we will derive a response time bound, IO-AMC-rtb, for Sporadic
Servers and PIBS.

\subsection{I/O Adaptive Mixed-Criticality Model}

Sporadic Servers follow a similar model to the original AMC model.  A Sporadic
Server task $\tau_i$ is assigned a criticality level $L_i {\in}
\left\{\mathtt{LO}, \mathtt{HI}\right\}$, a period $T_i$ and a vector of
capacities $\vec{C_i}$.  The deadline is assumed to be equal to the period.
If $L_i{=}\mathtt{LO}$, $\tau_i$ only runs while the system is in the
\LO{}-criticality mode and therefore only $C\left(\mathtt{LO}\right)$ is
defined.  For \HI{}-criticality tasks both $C\left(\mathtt{LO}\right)$ and
$C\left(\mathtt{HI}\right)$ are defined and $C\left(\mathtt{HI}\right) \geq
C\left(\mathtt{LO}\right)$.

For PIBS, an I/O task $\tau_k$ is again assigned to either the \LO{} or \HI{}
criticality level; $L_k {\in} \left\{\mathtt{LO}, \mathtt{HI}\right\}$.  As
previously discussed, PIBS are only defined by a utilization $U_k$.  The
period, deadline and priority for a PIBS is inherited from the Sporadic Server
for which it is performing a task.  For IO-AMC, this definition is extended
and each PIBS is defined by a vector of utilizations $\vec{U_k}$.  If $\tau_k$
is a \LO{}-criticality PIBS, i.e. $L_k {=} \mathtt{LO}$, then
${U_k}\left(\mathtt{LO}\right) {>} {U_k}\left(\mathtt{HI}\right)$ and
if $L_k {=} \mathtt{HI}$ then ${U_k} \left(\mathtt{LO}\right) {\leq}
{U_k}\left(\mathtt{HI}\right)$.  This definition allows \LO{}-criticality
PIBS to continue execution after the switch to \HI{}-criticality.  This model
allows users to assign criticality levels to I/O devices indirectly by
assigning criticality levels to the PIBS that execute in response to the I/O
device.

With the typical AMC model now augmented to consider PIBS we can now derive a
new admissions test for IO-AMC.  First, the PIBS interference equation
introduced in Section~\ref{sect:ss_and_pibs} is modified to incorporate
criticality levels:

\begin{align*}
I^q_k\left(t, L\right) = \left(1 + \left\lceil \frac{t}{T_q} \right\rceil -
U_k\left(L\right) \right) T_q U_k\left(L\right)
\end{align*}

As before, there are three conditions that must be considered: (1) the
\LO{}-criticality steady state, (2) the \HI{}-criticality steady state, and
(3) the change from \LO{}-criticality to \HI{}-criticality.  The steady states
are again simple and are merely extensions of the non-mixed-criticality
response time bounds.  For Sporadic Server tasks the steady state equations
are:
\begin{align}
  R^{\mathtt{LO}}_i =
  C_i\left(\mathtt{LO}\right)&+\sum_{\tau_j\in\mathbf{hp}\left(i\right)}
  \left\{\left\lceil\frac{R^{\mathtt{LO}}_i}{T_j}\right\rceil
    C_j\left(\mathtt{LO}\right)\right\} \notag \\ &+\sum_{\tau_k \in \mathbf{ps}}\max_{\tau_q \in
    \mathbf{hip}\left(i\right)} \left\{
    I^q_k\left(R^{\mathtt{LO}}_i, \mathtt{LO}\right) \right\} \\
  R^{\mathtt{HI}}_i =
  C_i\left(\mathtt{HI}\right)&+\sum_{\tau_j\in\mathbf{hpH}\left(i\right)}
  \left\{\left\lceil\frac{R^{\mathtt{HI}}_i}{T_j}\right\rceil
    C_j\left(\mathtt{HI}\right)\right\} \notag \\ &+\sum_{\tau_k \in \mathbf{ps}}\max_{\tau_q \in
    \mathbf{hipH}\left(i\right)} \left\{
    I^q_k\left(R^{\mathtt{HI}}_i, \mathtt{HI}\right) \right\}
\end{align}
where $\mathbf{hipH}\left(i\right){=}\mathbf{hpH}\left(i\right) {\cup}
\left\{\tau_i\right\}$, i.e. it is the set of all \HI{}-criticality tasks of
higher or equal priority than task $\tau_i$, plus task $\tau_i$ itself.  For
PIBS task $\tau_p$, running on behalf of Sporadic Server task $\tau_s$, the
steady state equations are:
\begin{align}
  _{s}R^{\mathtt{LO}}_p =
  &(2-U_p\left(\mathtt{LO}\right))U_p\left(\mathtt{LO}\right)T_s \notag \\
  &+\sum_{\tau_j\in\mathbf{hip}\left(s\right)}
  \left\{\left\lceil\frac{_{s}R^{\mathtt{LO}}_p}{T_j}\right\rceil
    C_j\left(\mathtt{LO}\right)\right\} \notag \\ &+\sum_{\tau_k \in
    \mathbf{ps}\setminus\left\{\tau_p\right\}}\max_{\tau_q \in
    \mathbf{hip}\left(s\right)} \left\{
    I^q_k\left(_{s}R^{\mathtt{LO}}_p, \mathtt{LO}\right) \right\}
\end{align}
\begin{align}
  _{s}R^{\mathtt{HI}}_p =
  &(2-U_p\left(\mathtt{HI}\right))U_p\left(\mathtt{HI}\right)T_s  \notag \\
  &+\sum_{\tau_j\in\mathbf{hipH}\left(s\right)}
  \left\{\left\lceil\frac{_{s}R_p^{\mathtt{HI}}}{T_j}\right\rceil
    C_j\left(\mathtt{HI}\right)\right\} \notag \\ &+\sum_{\tau_k \in
    \mathbf{ps} \setminus\left\{\tau_p\right\}}\max_{\tau_q \in
    \mathbf{hipH}\left(s\right)} \left\{
    I^q_k\left(_{s}R_p^{\mathtt{HI}}, \mathtt{HI}\right) \right\}
\end{align}
As with the traditional response time analysis of PIBS, its deadline is the same
as that of its corresponding Sporadic Server~$\tau_s$. Therefore, the above
analysis must be applied to all Sporadic Servers associated with a PIBS.

\subsection{IO-AMC-rtb}

The techniques described in Section~\ref{sect:amc} are used for the IO-AMC-rtb
analysis.  Specifically, \LO{}-criticality PIBS are allowed to continue
execution in the \HI{}-criticality mode.  For a Sporadic Server task the
IO-AMC-rtb equation is:
\begin{align} \label{eq:amc_ss_and_pibs_ss}
  R_i^* = C_i&+ \sum_{\tau_j \in
    \mathbf{hpH}\left(i\right)}{\left\lceil\frac{R_i^*}{T_j}\right\rceil
    C_j\left(\mathtt{HI}\right)} \notag \\ &+ \sum_{\tau_j \in
    \mathbf{hpL}\left(i\right)}{\left\lceil\frac{R_i^{\mathtt{LO}*}}{T_j}\right\rceil
    C_j\left(\mathtt{LO}\right)} \notag \\ &+
    \sum_{\tau_k\in\mathbf{psH}} \bigg\{ \max_{\tau_q \in \mathbf{hip}\left(i\right)}
      I^q_k\left(R_i^*, \mathtt{HI}\right) \bigg\}  \notag \\ &+
    \sum_{\tau_k\in\mathbf{psL}} \bigg\{\max_{\tau_q \in \mathbf{hip}\left(i\right)}
      I^q_k\left(R_i^{\mathtt{LO}*}, \mathtt{LO}\right) + \notag \\
      &\;\max_{\tau_{q^\prime} \in \mathbf{hipH}\left(i\right)}
      I^{q^\prime}_k\left(R_i^*-R_i^{\mathtt{LO}*}, \mathtt{HI}\right)
  \bigg\}
\end{align}
where $\mathbf{psH}$ and $\mathbf{psL}$ are the set of \HI{} and
\LO{}-criticality PIBS respectively.  The last summation in
Equation~\ref{eq:amc_ss_and_pibs_ss} represents the maximum interference a
\LO{}-criticality PIBS can cause.  Specifically,
$I^q_k\left(R_i^\mathtt{LO}, \mathtt{LO}\right)$ represents the maximum
interference the PIBS can cause before the mode change and
$I^{q^\prime}_k\left(R_i^*-R_i^{\mathtt{LO}}, \mathtt{HI}\right)$ represents the
total interference the PIBS can cause after the mode change.  Again, the
Sporadic Server that maximizes the interference is chosen for each PIBS.

The IO-AMC-rtb equation for a PIBS $\tau_k$ when assigned to Sporadic Server
$\tau_s$ is:
\begin{align} \label{eq:amc_ss_and_pibs_pibs}
  _{s}R_p^{*} = &(2-U_p\left(\mathtt{HI}\right))T_sU_p\left(\mathtt{HI}\right) \notag \\ & +
  \sum_{\tau_j \in \mathbf{hipH}\left(s\right)}{\left\lceil\frac{_{s}R_p^*}{T_j}\right\rceil
    C_j\left(\mathtt{HI}\right)} \notag \\ &+ \sum_{\tau_j \in
    \mathbf{hipL}\left(s\right)}{\left\lceil\frac{_{s}R_p^{\mathtt{LO}*}}{T_j}\right\rceil
    C_j\left(\mathtt{LO}\right)} \notag \\ &+
    \sum_{\tau_k\in\left(\mathbf{psH} \setminus \left\{\tau_p\right\}\right)}  \bigg\{\max_{\tau_q \in \mathbf{hip}\left(s\right)}
      I^q_k\left(_{s}R_p^*, \mathtt{HI}\right) \bigg\}  \notag \\ &+
    \sum_{\tau_k\in\left(\mathbf{psL}\setminus \left\{\tau_p\right\}\right)} \bigg\{\max_{\tau_q \in \mathbf{hip}\left(s\right)}
      I^q_k\left(_{s}R_p^{\mathtt{LO}*}, \mathtt{LO}\right) + \notag \\
      &\;\max_{\tau_{q^\prime} \in \mathbf{hipH}\left(s\right)}\
      I^{q^\prime}_k\left(_{s}R_p^* - {_{s}R_p^{\mathtt{LO}*}}, \mathtt{HI}\right)
  \bigg\}
\end{align}
Equation~\ref{eq:amc_ss_and_pibs_pibs} differs from
Equation~\ref{eq:amc_ss_and_pibs_ss} in the first term, and by the exclusion
of $\tau_p$ from the set of PIBS.  Similar to
Equation~\ref{eq:pibs_rta}, the response time analysis requires iterating over
all \HI{}-criticality Sporadic Servers that could be associated with
the PIBS.  This is because only the \HI{}-criticality Sporadic Servers are of
interest after the mode change.

As mentioned in Section~\ref{sect:amc}, recent related work by Burns and
Baruah~\cite{BurnsBa13} has extended the original AMC model to allow
\LO{}-criticality tasks to continue running after the mode to the
\HI{}-criticality mode.  The derivation this work provided was used to allow
\LO{}-criticality PIBS to continue running after the mode change.  This work can
also be applied to allow \LO{}-criticality Sporadic Servers to continue running
in the \HI{}-criticality mode.  This derivation is similar to the one provided
by Burns and Baruah but there are subtle differences due to the inclusion of
PIBS.

First for Sporadic Servers during the mode change the new IO-AMC-rtb equation
is:
\begin{align} \label{eq:amc_ss_and_pibs_ss_extended}
  R_i^* = C_i&+ \sum_{\tau_j \in
    \mathbf{hpH}\left(i\right)}{\left\lceil\frac{R_i^*}{T_j}\right\rceil
    C_j\left(\mathtt{HI}\right)} \notag \\ &+ \sum_{\tau_j \in
    \mathbf{hpL}\left(i\right)}{\left\lceil\frac{R_i^{\mathtt{LO}*}}{T_j}\right\rceil
    C_j\left(\mathtt{LO}\right)} \notag \\ &+\sum_{\tau_j \in
    \mathbf{hpL}\left(i\right)}{\left(\left\lceil\frac{R_i^*}{T_j}\right\rceil -
      \left\lceil\frac{R_i^{\mathtt{LO}*}}{T_j}\right\rceil\right)
    C_j\left(\mathtt{HI}\right)} \notag \\ &+
    \sum_{\tau_k\in\mathbf{psH}} \bigg\{ \max_{\tau_q \in \mathbf{hip}\left(i\right)}
      I^q_k\left(R_i^*, \mathtt{HI}\right) \bigg\}  \notag \\ &+
    \sum_{\tau_k\in\mathbf{psL}} \bigg\{\max_{\tau_q \in \mathbf{hip}\left(i\right)}
      I^q_k\left(R_i^{\mathtt{LO}*}, \mathtt{LO}\right) + \notag \\
      &\;\max_{\tau_{q^\prime} \in \mathbf{hip}\left(i\right)}
      I^{q^\prime}_k\left(R_i^*-R_i^{\mathtt{LO}*}, \mathtt{HI}\right)
  \bigg\}
\end{align}

In addition to the third term which is taken from
Equation~\ref{eq:lower_tasks_all}, Equation~\ref{eq:amc_ss_and_pibs_ss_extended}
differs from Equation~\ref{eq:amc_ss_and_pibs_ss} (where \LO{}-criticality
Sporadic Servers do not run in the \HI{}-criticality mode) in that all Sporadic
Servers of higher or equal priority must be considered when accounting for the
interference from \LO{}-criticality PIBS after the mode change.  Specifically
the $\mathbf{hipH}$ in the final term has changed to $\mathbf{hip}$ to reflect
that fact that Sporadic Servers of all criticality levels run in the
\HI{}-criticality mode.

Equation~\ref{eq:amc_ss_and_pibs_pibs_extended} is the IO-AMC-rtb equation for
PIBS when \LO{}-criticality Sporadic Servers are allowed to run in the
\HI{}-criticality mode.  Again the only differences are the inclusion of the
interference caused by \LO{}-criticality tasks after the mode change and
changing the $\mathbf{hipH}$ to $\mathbf{hip}$ in the final term to account for
the fact that all Sporadic Servers are capable of executing after the mode
change.
\begin{align} \label{eq:amc_ss_and_pibs_pibs_extended}
  _{s}R_p^{*} = &(2-U_p\left(\mathtt{HI}\right))T_sU_p\left(\mathtt{HI}\right) \notag \\ & +
  \sum_{\tau_j \in \mathbf{hipH}\left(s\right)}{\left\lceil\frac{_{s}R_p^*}{T_j}\right\rceil
    C_j\left(\mathtt{HI}\right)} \notag \\ &+ \sum_{\tau_j \in
    \mathbf{hipL}\left(s\right)}{\left\lceil\frac{_{s}R_p^{\mathtt{LO}*}}{T_j}\right\rceil
    C_j\left(\mathtt{LO}\right)} \notag \\ &+\sum_{\tau_j \in
    \mathbf{hpL}\left(i\right)}{\left(\left\lceil\frac{_{s}R_p^*}{T_j}\right\rceil -
      \left\lceil\frac{_{s}R_p^{\mathtt{LO}*}}{T_j}\right\rceil\right)
    C_j\left(\mathtt{HI}\right)} \notag \\ &+
    \sum_{\tau_k\in\left(\mathbf{psH} \setminus \left\{\tau_p\right\}\right)}  \bigg\{\max_{\tau_q \in \mathbf{hip}\left(s\right)}
      I^q_k\left(_{s}R_p^*, \mathtt{HI}\right) \bigg\}  \notag \\ &+
    \sum_{\tau_k\in\left(\mathbf{psL}\setminus \left\{\tau_p\right\}\right)} \bigg\{\max_{\tau_q \in \mathbf{hip}\left(s\right)}
      I^q_k\left(_{s}R_p^{\mathtt{LO}*}, \mathtt{LO}\right) + \notag \\
      &\;\max_{\tau_{q^\prime} \in \mathbf{hip}\left(s\right)}\
      I^{q^\prime}_k\left(_{s}R_p^* - {_{s}R_p^{\mathtt{LO}*}}, \mathtt{HI}\right)
  \bigg\}
\end{align}

\section{Implementation}
\label{sect:implementation}
Changes to the existing Quest scheduler to support IO-AMC scheduling include:

\begin{list}
{$\bullet$}
{
\setlength{\topsep}{\parskip}
\setlength{\parsep}{0in}
\setlength{\itemsep}{0in}
\setlength{\parskip}{0in}
}
\item Mapping Quest processes to a task and job model;
\item Detecting when to change to \HI{}-criticality mode;
\item Adjusting Sporadic Server and PIBS replenishments.
\end{list}

Quest tasks are assigned to Sporadic Servers, and are similar
to UNIX processes in that they run until an \texttt{\_exit} system call is
invoked.  In comparison, real-time tasks in the IO-AMC model release a job at
a specific rate and the job runs until completion.  To accommodate the
differences, a new \texttt{sync} system call was introduced.  The
\texttt{sync} function indicates the start and end of real-time jobs.  A
typical IO-AMC task has a setup phase, followed by a loop which repeatedly
calls \texttt{sync} before starting the body of the next job.

A typical IO-AMC application will look similar to the procedure outlined in
Algorithm~\ref{alg:amc-application}.

\begin{algorithm}
\caption{IO-AMC User Space Process}
\label{alg:amc-application}
\footnotesize
\begin{algorithmic}
  \Procedure {main}{}
  \State // Setup Code
  \While {$\mathtt{TRUE}$}
    \State $\mathtt{sync\left(\right)}$
    \State // Job Code
  \EndWhile
  \EndProcedure
\end{algorithmic}
\end{algorithm}

An alternative approach would be to have one process repeatedly fork another
process, with the newly forked process being the job for that period.  This
approach would not require the application developer to invoke the \texttt{sync}
system call as the \texttt{\_exit} system call could be used to inform the
scheduler that the job is completed.  This approach was not chosen however due
to the extra overhead involved in the creation of a new process.

The first time \texttt{sync} is called, a mixed-criticality task will sleep
until its Sporadic Server is replenished to full capacity. The deadline will
be set $T$ time units after the process wakes up.  Subsequent calls to
\texttt{sync} will have the process sleep up to the deadline, emulating the
job being completed and waiting for the next job.

Quest bottom half threads are assigned to Priority Inheritance Bandwidth
Preserving Servers, and follow the real-time task and
job model.  Bottom half threads are only woken up by the top halves, and at
the end of their execution they notify the scheduler of their
completion. Therefore, no changes needed to be made to the PIBS and bottom
half threads.

The conditions for a mode change depend on whether a task is running a job or
bottom half on a Sporadic Server or PIBS.  A \HI{}-criticality Sporadic Server
initiates a mode change when it has depleted all replenishment items that are
before the deadline.  If this happens the Sporadic Server will not be able to
run until after the deadline and therefore a mode change must occur.  A
\HI{}-criticality PIBS causes a mode change when it has depleted its budget
before completing the bottom half thread.  In this case, the single
replenishment for a PIBS will be at the deadline and therefore the PIBS
will not be able to run until after its deadline unless a mode change
occurred.

Finally, when the mode change occurs, the Sporadic Server and PIBS
replenishment items must be adjusted to take into account the new or removed
budget.  For \HI{}-criticality Sporadic Servers, the difference between
$C\left(\texttt{HI}\right)$ and $C\left(\texttt{LO}\right)$ is added to the
beginning of the first replenishment list item, if the item's replenishment
time is equal to or less than the current time, or if the replenishment list
is full.  Otherwise, a new replenishment list item is inserted at the
beginning with a replenishment time equal to the current time.  A
replenishment item $R$ has two properties, $R.amt$, which is the amount of
budget replenished, and $R.time$, which is when the replenishment occurs.
This is demonstrated in Algorithm~\ref{alg:HI-SS-adjustment}.

\begin{algorithm}
\caption{\texttt{HI}-Criticality Sporadic Server Adjustment}
\label{alg:HI-SS-adjustment}
\footnotesize
\begin{algorithmic}
  \State \Comment {$S$ is the Sporadic Server being modified}
  \State $now \gets$ current time
  \State $additional\_cap \gets C\left(\mathtt{HI}\right) - C\left(\mathtt{LO}\right)$
  \If {$\left(S.Q.head.time \leq now\right) \mathtt{OR} \left(\mathtt{MAX\_LENGTH} = S.Q.length\right)$}
    \State $S.Q.head.amt \gets S.Q.head.amt + additional\_cap$
  \Else
    \State \Comment{$R$ is a new replenishment item}
    \State $R.time \gets now$
    \State $R.amt \gets additional\_cap$
    \State $S.Q.add\left(R\right)$
  \EndIf
\end{algorithmic}
\end{algorithm}

For \LO{}-criticality Sporadic Servers, the adjustment algorithm is more
complicated.  
If $C\left(\mathtt{LO}\right)
{-} C\left(\mathtt{HI}\right)$ is less than or equal to the remaining budget
for this period, i.e. before the deadline, then budget is removed from the
replenishment list by moving backwards in time from the replenishment item
right before the deadline. The head replenishment item must be treated
differently if the Sporadic Server has used some of that budget at the time of
a mode change. $S.usage$ tracks how much has been used from the head
replenishment item.  If $S.usage$ is greater than zero, a new replenishment
item might be posted at the end of the queue with an amount equal to
$S.usage$.  If $C\left(\mathtt{LO}\right) {-} C\left(\mathtt{HI}\right)$ is
greater than the remaining budget for this period then the Sporadic Server has
run for more than its $C\left(\mathtt{HI}\right)$ for this period.  In this
case all the available budget in the replenishment list is removed for this
period and the difference is removed from the end of the replenishment list.
Algorithm~\ref{alg:LO-SS-adjustment} contains the pseudocode for adjusting the
budget of a \LO{}-criticality Sporadic Server.

\begin{algorithm}
\caption{\texttt{LO}-Criticality Sporadic Server Adjustment}
\label{alg:LO-SS-adjustment}
\footnotesize
\begin{algorithmic}
  \State \Comment {$S$ is the Sporadic Server being modified}
  \State $R_d \gets$ replenishment item in $S.Q$ right before the deadline of $S$
  \State \Comment{$R_d$ is $NULL$ if no such replenishment item exists}
  \State $now \gets$ current time
  \State $reduced\_cap \gets C\left(\mathtt{LO}\right) - C\left(\mathtt{HI}\right)$
  \While {$\left(reduced\_cap > 0\right) \mathtt{AND} \left(R_d \neq NULL\right)$}
    \If {$\left(R_d = S.Q.head\right) \mathtt{AND} \left(S.usage > 0\right)$}
      \If {$R_d.amt - S.usage > reduced\_cap$}
        \State $R_d.amt \gets R_d.amt - reduced\_cap$
        \State $reduced\_cap \gets 0$
      \Else
        \State $reduced\_cap \gets reduced\_cap - \left(R_d.amt - S.usage\right)$
        \State $S.Q.remove\left(R_d\right)$
        \State $R_d.amt \gets S.usage$
        \State $R_d.time \gets R_d.time + S.period$
        \State $S.Q.add\left(R_d\right)$
        \State $S.usage \gets 0$
      \EndIf
      \State $R_d \gets NULL$
    \Else
      \If {$R_d.amt \leq reduced\_cap$}
        \State $reduced\_cap \gets reduced\_cap - R_d.amt$
        \State $R_{tmp} \gets R_d.prev$
        \State $S.Q.remove\left(R_d\right)$
        \State $R_d \gets R_{tmp}$
      \Else
        \State $R_d.amt \gets R_d.amt - reduced\_cap$
        \State $reduced\_cap \gets 0$
      \EndIf
    \EndIf
  \EndWhile
  \While {$reduced\_cap > 0$}
    \If {$S.Q.end.amt \leq reduced\_cap$}
      \State $reduced\_cap \gets reduced\_cap - S.Q.end.amt$
      \State $S.Q.remove\left(S.Q.end\right)$
    \Else
        \State $S.Q.end.amt \gets S.Q.end.amt - reduced\_cap$
        \State $reduced\_cap \gets 0$
    \EndIf
  \EndWhile
\end{algorithmic}
\end{algorithm}

PIBS have a much simpler mode change algorithm because there is only one
replenishment item to consider for each invocation of the server.  Also, due
to the aperiodicity of I/O events, a deadline for a PIBS is calculated
from when an I/O event is initiated.  This results in pessimistic analysis in
Equations~\ref{eq:amc_ss_and_pibs_ss} and~\ref{eq:amc_ss_and_pibs_pibs}.
We have to assume that regardless of whether a PIBS is \LO{} or
\HI{}-criticality, it causes the maximum interference possible in both the
\LO{} and \HI{} modes.  Therefore, both \LO{} and
\HI{} PIBS can simply replenish their full budget at the time of a mode
change.

\section{Evaluation}
\label{sect:experimental}
The experimental evaluation consists of two sections: 1) simulation based
schedulability tests and 2) experiments conducted using the IO-AMC
implementation in the Quest operating system.  The simulation based
schedulability tests show that a system of Sporadic Servers and PIBS have a
similar but slightly lower schedulability than a system of just Sporadic
Servers.  This is due to the extra interference that can be caused by PIBS
compared to Sporadic Servers.  The Quest experiments show the benefits of PIBS
compared to Sporadic Servers and how mixed-criticality can be used to control
the bandwidth from I/O devices with different criticalities.

\subsection{Simulation Experiments}

In order to compare the proposed scheduling approaches, random task sets were
generated with varying total utilizations.  500 task sets were generated for
each utilization value ranging from $0.20$ to $0.95$ with $0.05$ increments.
Each task set was tested to see if it was schedulable under the different
policies.  Each PIBS was randomly assigned to a single Sporadic Server of the
same criticality level.  For systems comprising only Sporadic Servers, the
PIBS were converted to Sporadic Servers of equivalent utilization and
period.\footnote{The PIBS period was set equal to its corresponding Sporadic
  Server.} The parameters used to generate the task sets used are outlined in
Table~\ref{table:sim-parameters}.

\begin{table}[h!]
  \centering
  \begin{tabular}{|c|c|}  \hline
    \textbf{Parameter} & \textbf{Value} \\ \hline
    Number of Tasks & 20 (15 Main, 5 I/O)  \\ \hline
    Criticality Factor & 2  \\ \hline
    Probability $L_{i}=\mathtt{HI}$ & 0.5  \\ \hline
    Period Range & 1 -- 100  \\ \hline
    I/O Total Utilization & 0.05  \\ \hline
  \end{tabular}
\caption{Parameters Used to Generate Task Sets}
\label{table:sim-parameters}
\end{table}
The UUnifast algorithm~\cite{BiniBu05} was used to generate the task
sets. Task periods were generated with a log-uniform distribution.  For the
mixed-criticality experiments, $C_{i}(\mathtt{LO}){=}U_{i}/T_{i}$. If
$L_{i}{=}\mathtt{HI}$, $C_{i}(\mathtt{HI}){=}\mathtt{CF} \times
C_{i}(\mathtt{LO})$, where $\mathtt{CF}$ is the criticality factor. For our
experiments, if $L_{i}{=}\mathtt{LO}$, $C_{i}(\mathtt{HI}){=}0$.

The following are the different types of schedulability tests that were used
in the evaluation.  This includes schedulability tests for mixed-criticality
and traditional systems.
\begin{itemize}[leftmargin=*]

\item \textbf{SS-rta} -- Sporadic Server response time analysis.  Due to the
  nature of Sporadic Servers, this is the same as a periodic response time
  analysis.

\item \textbf{SS+PIBS-rta} -- Sporadic Server and PIBS response time analysis
  introduced in this paper.  See Section~\ref{sect:ss_and_pibs}.

\item \textbf{AMC-rtb} -- Adaptive Mixed-Criticality response time bound
  developed by Baruah et al.~\cite{BaruahBuDa11}. See Section~\ref{sect:amc}.

\item \textbf{IO-AMC-rtb} -- I/O Adaptive Mixed-Criticality response time bound
  developed in this paper.  See Section~\ref{sect:amc_ss_and_pibs}.

\item \textbf{AMC UB} -- This is not a schedulability test but instead an
  upper bound for AMC.  It consists of both the \LO{}- and \HI{}-criticality
  level steady states tests. See Section~\ref{sect:amc} for details.

\item \textbf{IO-AMC UB} -- This is not a schedulability test but instead
  an upper bound for IO-AMC.  It consists of both the \LO{}- and
  \HI{}-criticality level steady states tests.  See
  Section~\ref{sect:amc_ss_and_pibs} for details.

\end{itemize}

\subsubsection{SS+PIBS vs. SS-Only Simulations}

\begin{figure}[ht]
  \centering
  \includegraphics[width=0.5\textwidth]{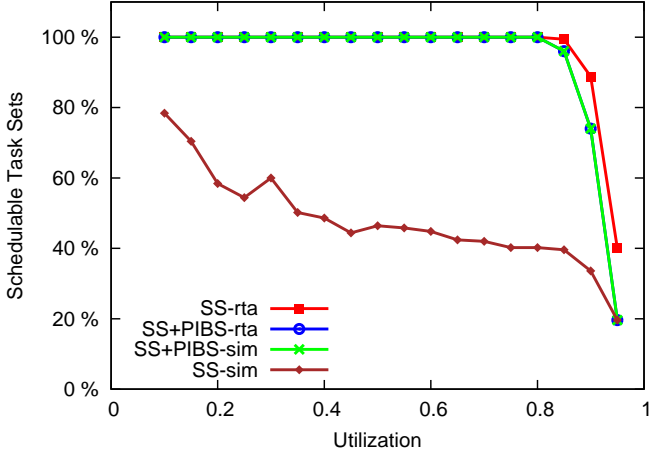}
  \caption{Schedulability of SS+PIBS vs SS-Only}
  \label{fig:rta}
\end{figure}

Figure~\ref{fig:rta} contains the results of the response time analysis and
event simulator for a system of Sporadic Servers and PIBS (SS+PIBS) compared
to a system of only Sporadic Servers (SS-Only).  As expected, a higher number
of the Sporadic Server only task sets are schedulable using the response time
analysis equations compared to the SS+PIBS response time analysis.  This is
due to the extra interference a PIBS can cause compared to a Sporadic Server
of equivalent utilization and period.

\subsubsection{IO-AMC vs. AMC Simulations}

In this section, IO-AMC is compared to an AMC system containing only
Sporadic Servers under different mixed-criticality scenarios.

\begin{figure}[h!]
  \centering
  \includegraphics[width=0.5\textwidth]{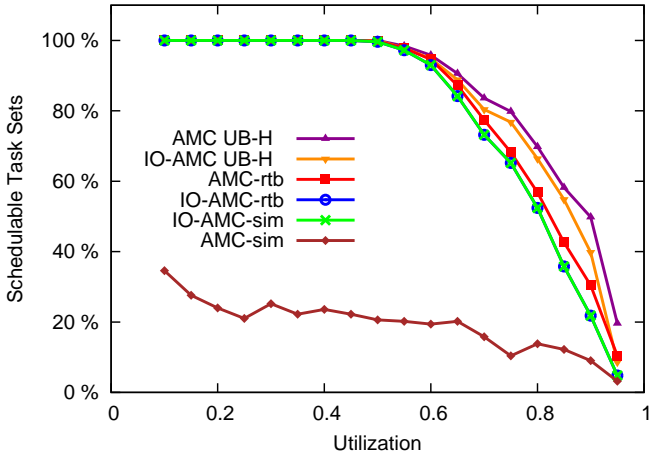}
  \caption{Schedulability of IO-AMC vs AMC}
  \label{fig:amc-rtb}
\end{figure}

Figure~\ref{fig:amc-rtb} shows the response time analysis and simulation
results when \LO{}-criticality tasks do not run in the \HI{}-criticality mode.
Similar to Figure~\ref{fig:rta}, AMC-rtb outperforms
IO-AMC-rtb.  This is due to the fact that AMC-rtb is an
extension of the traditional response time analysis and does not experience
the extra interference caused by PIBS. 

We also varied task set parameters to identify their effects on schedulability. 
For each set of parameters $p$ in a given test $y$, we measured the weighted
schedulability~\cite{BastoniBrAn10}, which is defined as follows:
\[
W_y\left(p\right){=}\sum_{\forall \tau} \left(u\left(\tau\right) \times
  S_y\left(\tau, p\right)\right) / \sum_{\forall \tau} u\left(\tau\right)
\]
where $S_y\left(\tau, p\right)$ is the binary result (0 or 1) of the
schedulability test $y$ on task set $\tau$, and $u(\tau)$ is the total
utilization.  The weighted schedulability compresses a three-dimensional plot to
two dimensions and places higher value on task sets with higher utilization.

\begin{figure}[!ht]
  \vspace{-0.05in}
  \includegraphics[width=0.5\textwidth]{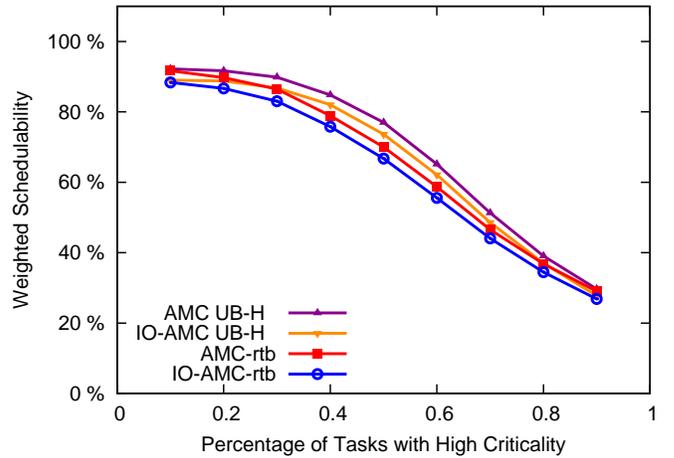}
  \vspace{0.01in}
  \caption{Weighted Schedulability vs \% of \HI{}-criticality Tasks}
  \label{fig:amc-perc}
\end{figure}

\begin{figure}[!ht]
  \vspace{-0.07in}
  \includegraphics[width=0.5\textwidth]{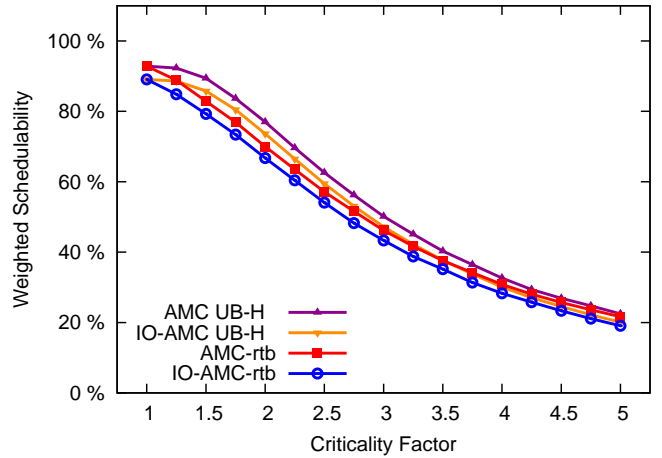}
  \vspace{0.01in}
  \caption{Weighted Schedulability vs Criticality Factor}
  \label{fig:amc-factor}
\end{figure}

\begin{figure}[!ht]
  \vspace{-0.07in}
  \includegraphics[width=0.5\textwidth]{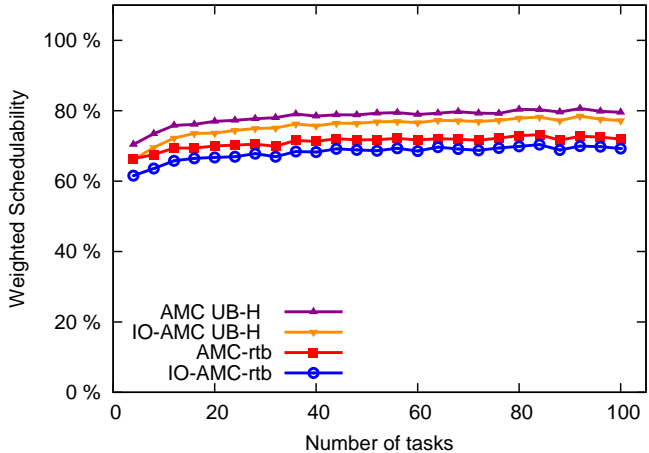}
  \vspace{0.01in}
  \caption{Weighted Schedulability vs Number of Tasks}
  \label{fig:amc-tasks}
\end{figure}

Figures~\ref{fig:amc-perc}, \ref{fig:amc-factor}, and \ref{fig:amc-tasks} show
the results of varying the probability of a \HI{}-criticality task, the
criticality factor and the number of tasks respectively.  In all
scenarios, \LO{}-criticality tasks do not run in the \HI{}-criticality mode.  As
expected, the percentage of schedulable tasks for IO-AMC is slightly lower than
the percentage for traditional AMC.  This is again due to the slightly larger
interference caused by a PIBS.

\subsection{Quest Experiments}

The above simulation results do not capture the practical costs of a system of
servers for tasks and interrupt bottom halves. This section investigates the
performance of our IO-AMC policy in the Quest real-time system. We also study
the effects of mode changes on I/O throughput for an application that
collects streaming camera data. All experiments were run on a 3.10 GHz
Intel\textsuperscript{\textregistered} Core i3-2100 CPU.

\subsubsection{Scheduling Overhead}

We studied the scheduling overheads for two different system implementations
in Quest. In the first system, Sporadic Servers were used for both tasks and
bottom halves (SS-Only). In the second system, Sporadic Servers were used for
tasks, and PIBS were used to handle interrupt bottom halves (SS+PIBS). In both
cases, a task set consisted of two application threads of different
criticality levels assigned to two different Sporadic Servers, and one bottom
half handler for interrupts from a USB camera. The first application thread
read all the data available from the camera in a non-blocking manner and then
busy-waited for its entire budget to simulate the time to process the
data. The second application thread simply busy-waited for its entire budget,
to simulate a CPU-bound task without any I/O requests. Both application
threads consisted of a sequence of jobs. Each job was released once every
server period or immediately after the completion of the previous job,
depending on which was later. The experimental parameters are shown in
Table~\ref{tbl:servers}.

\begin{table}[ht]
\begin{tabular}{  l  c  c  r  }
  Task & $C\left(\mathtt{LO}\right)$ or $U\left(\mathtt{LO}\right)$ & $C\left(\mathtt{HI}\right)$ or $U\left(\mathtt{HI}\right)$ & $T$ \\
  \hline
  Application 1\\(\HI{}-criticality) & $23{ms}$ & $40{ms}$ & $100{ms}$ \\
  \hline
  Application 2\\(\LO{}-criticality) & $10{ms}$ & $1{ms}$ & $100{ms}$ \\
  \hline
  Bottom Half (PIBS)& $U\left(\mathtt{LO}\right)=1\%$ & $U\left(\mathtt{HI}\right)=2\%$ & $100{ms}$ \\
  \hline
  Bottom Half (SS)& $1{ms}$ & $2{ms}$ & $100{ms}$ \\
  \hline
\end{tabular}
\caption{Quest Task Set Parameters for Scheduling Overhead}
\label{tbl:servers}
\end{table}

\begin{figure}[ht]
  \centering
  \includegraphics[width=0.5\textwidth]{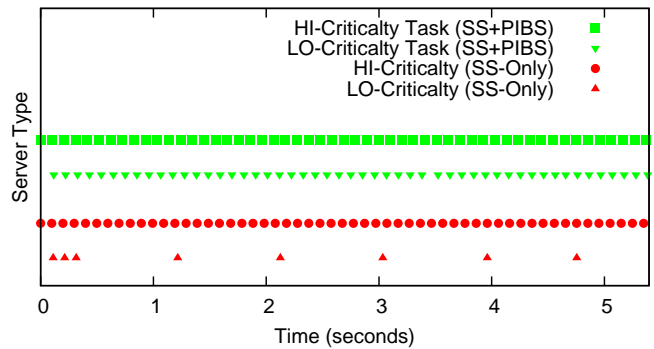}
  \caption{Job Completion Times for SS+PIBS vs SS-Only}
  \label{fig:interference}
\end{figure}

The processor's timestamp counter was recorded when each application finished
its current job. Results are shown in Figure~\ref{fig:interference}. For
SS+PIBS, each application completed its jobs at regular intervals.  However,
for SS-Only, the \HI{}-criticality server for interrupts from the USB camera
caused interference with the application tasks. This led to the
\HI{}-criticality task depleting its budget before finishing its job.  This is
due to the extra overhead added by a Sporadic Server handling the interrupt
bottom half thread.  Therefore, the system had to switch into the
\HI{}-criticality mode to ensure the \HI{}-criticality task completed its job,
sacrificing the performance of the \LO{}-criticality task.  This is depicted
by the larger time between completed jobs in Figure~\ref{fig:interference}.
The SS+PIBS task set did not suffer from this problem due to the lower
scheduling overhead caused by PIBS.

\begin{figure}[ht]
  \centering
  \includegraphics[width=0.5\textwidth]{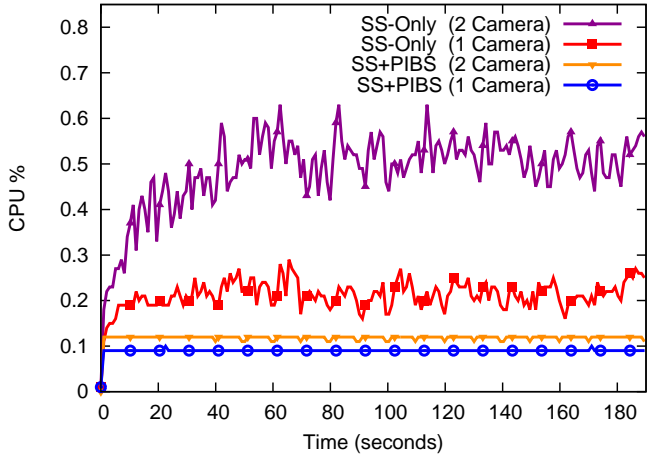}
  \caption{Scheduling Overheads for SS-Only vs SS+PIBS}
  \label{fig:overhead}
\end{figure}

Figure~\ref{fig:overhead} shows the additional overhead caused when Sporadic
Servers are used for bottom half threads as opposed to PIBS.  This higher
scheduling overhead is the cause for the mode change in the previous
experiment.  Figure~\ref{fig:overhead} depicts two different system
configurations, one involving only a single camera and another involving two
cameras. For each configuration, the scheduling overhead for both SS-Only and
SS+PIBS was measured.  For the single camera configuration, there is one
\HI{}-criticality task, one \LO{}-criticality task, and one \HI{}-criticality
server (either PIBS or Sporadic Server) for the USB camera interrupt bottom
half thread.  The scheduling overhead for SS-Only is more erratic and higher
than the system of Sporadic Servers and PIBS.  The second configuration adds a
\LO{}-criticality camera with a $2\%$ utilization in the \LO{}-criticality
mode, a $1\%$ utilization in the \HI{}-criticality mode, and a period of $100$
microseconds when utilizing a Sporadic Server.  Figure~\ref{fig:overhead}
shows that the scheduling overhead for an SS-Only system more than doubled,
going from an average of $0.21\%$ to $0.49\%$, while an SS+PIBS system
experienced only a small increase of $0.03\%$.

\subsubsection{Mode Change for I/O Device}

As mentioned in Section~\ref{sect:amc_ss_and_pibs}, assigning criticality levels
to bottom half interrupt handlers is akin to assigning criticality levels to the
device associated with the bottom half.  To test this assertion, two USB
cameras were assigned different criticality levels and a mode change was caused
during the execution of the task set.  The task set consisted of two Sporadic
Servers and two PIBS, as shown in Table~\ref{tbl:2camera-servers}.

\begin{table}[ht]
  \vspace{0.1in}
\begin{tabular}{  l  c  c  r  }
  Task & $C\left(\mathtt{LO}\right)$ or $U\left(\mathtt{LO}\right)$ & $C\left(\mathtt{HI}\right)$ or $U\left(\mathtt{HI}\right)$ & $T$ \\
  \hline
  Application 1\\(\HI{}-criticality) & $25{ms}$ & $40{ms}$ & $100{ms}$ \\
  \hline
  Application 2\\(\LO{}-criticality) & $25{ms}$ & $24{ms}$ & $100{ms}$ \\
  \hline
  Camera 1 -- PIBS\\(\HI{}-criticality)& $U\left(\mathtt{LO}\right)=0.1\%$ & $U\left(\mathtt{HI}\right)=1\%$ & $100{ms}$ \\
  \hline
  Camera 2 -- PIBS\\(\LO{}-criticality)& $U\left(\mathtt{LO}\right)=1\%$ & $U\left(\mathtt{HI}\right)=0.1\%$ & $100{ms}$ \\
  \hline
\end{tabular}
\caption{Quest Task Set Parameters for I/O Device Mode Change}
\label{tbl:2camera-servers}
\end{table}

\begin{figure}[ht]
  \centering
  \includegraphics[width=0.5\textwidth]{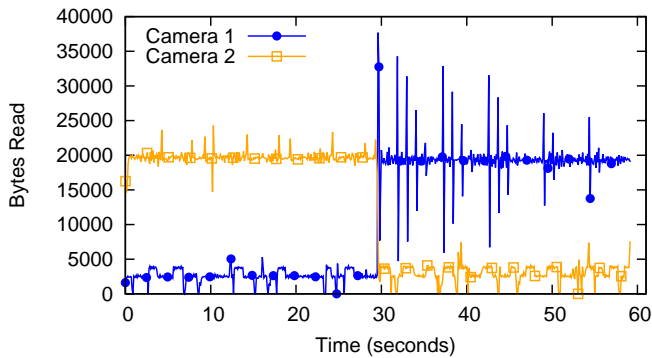}
  \caption{Data From \HI{}- and \LO{}-criticality USB Cameras}
  \label{fig:2camera-bandwidth}
  \vspace{-0.2in}
\end{figure}

\begin{figure}[ht]
  \centering
  \includegraphics[width=0.5\textwidth]{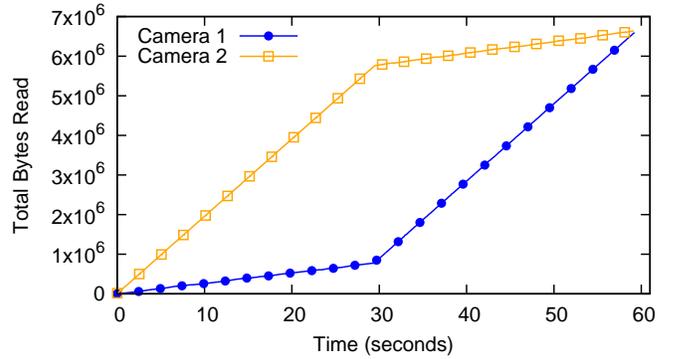}
  \caption{Total Data Processed Over Time}
  \label{fig:2camera-bandwidth-full}
  \vspace{-0.2in}
\end{figure}

Figure~\ref{fig:2camera-bandwidth} shows the camera data available at each
data point.  At approximately 30 seconds, a mode change occurs that causes
Camera~1 to change from a utilization of $0.1\%$ to $1\%$, thereby increasing
the amount of data received. Also at the time of the mode change, Camera~2's
utilization switches from $1\%$ to $0.1\%$, causing a drop in received
data. The variance for Camera~1 after the mode change is due to extra
processing of the delayed data that is performed by the bottom half interrupt
handler. Finally, Figure~\ref{fig:2camera-bandwidth-full} shows the total data
processed from each camera over time.

\section{Related Work}
\label{sect:related}
The most relevant related work has already been mentioned in
Sections~\ref{sect:ss_and_pibs} and~\ref{sect:amc}.  This section discusses
additional work in the areas of mixed-criticality and I/O-aware scheduling.

In recent years, there have been many extensions to the Adaptive
Mixed-Criticality model.  For single core scheduling, Barauh, Burns and Davis
extended their original AMC model to allow priorities to
change~\cite{BaruahBuDa13}.  Burns and Davis also introduced AMC-NPR
(Non-Preemptive Region), which improved schedulability by permitting tasks to
have a final non-preemptive region at the end of a job~\cite{BurnsDa14}.
Fleming and Burns extended the AMC model to allow more than two criticality
levels~\cite{FlemingBu13}.  These variations on the mixed-critical model could
be incorporated into the IO-AMC model.

Li and Barauh~\cite{LiBa12} combined the EDF-VD~\cite{BaruahBoDa11}
single-core mixed-criticality approach with fpEDF~\cite{Baruah04}, to develop
a multi-core mixed-criticality scheduling algorithm. Pathan also developed a
multi-core fixed priority scheduling algorithm for
mixed-criticality~\cite{Pathan12}. The approach adapted the original
single-core AMC approach to a multi-core scheduling framework compatible with
Audsley's algorithm~\cite{Audsley01}.  The work by Pathan is more likely to be
easily incorporated with the IO-AMC model given that both approaches use
fixed-priorities.

This work addresses the scheduling of and accounting of I/O events.  Lewandowski
et al.~\cite{LewandowskiStBa07} investigated the use of sporadic servers to
appropriately budget bottom half threads, as part of an Ethernet NIC device
driver.  Zhang and West developed a process-aware interrupt scheduling and
accountability scheme in Linux, to integrate the management of tasks and I/O
events~\cite{ZhangWe06}.  A similar approach was also implemented in the LITMUS
kernel for GPGPUs on multiprocessor systems~\cite{ElliottAn12}.

\section{Conclusions}
\label{sect:conclusions}
This paper builds on our scheduling framework in the Quest real-time operating
system, comprising a collection of Sporadic Servers for tasks and Priority
Inheritance Bandwidth-Preserving Servers (PIBS) for interrupt handlers. We
first show a response time analysis for a collection of Sporadic Servers and
PIBS in a system without mixed criticality levels. We then extend the analysis
to support an I/O Adaptive Mixed-Criticality (IO-AMC) model in a system
comprising of tasks and interrupt handlers. Our IO-AMC response time bound
considers a mode change to high-criticality when insufficient resources exist
for either high-criticality tasks or interrupt handlers in low-criticality
mode. The analysis considers the interference from low-criticality tasks and
interrupt handlers before the mode change.

Simulation results show that a system of only Sporadic Servers for both tasks
and interrupt handlers has a higher theoretical number of schedulable task
sets. However, in practice, using PIBS to handle interrupts is shown to be
superior because of lower system overheads. This paper also shows experimental
results in the Quest real-time operating system, where criticality levels are
assigned to devices. This enables high criticality devices to gain more
computational time when insufficient resources exist to service both high and
low criticality tasks and interrupt bottom halves. In turn, this enables high
criticality tasks that issue I/O requests to be granted more CPU time to meet
their deadlines.

The analysis in this paper assumes that tasks and I/O bottom half interrupt
handlers are executed on separate servers that are independent of one another.
In practice, a task may be blocked from execution until a pending I/O request
is completed. As long as the I/O request is handled within the time that a
task is waiting for its server to have its budget replenished, and is
therefore ineligible to run, then our analysis holds. Future work will
consider more complex task models where I/O requests can lead to blocking
delays that impact the execution of tasks.

\bibliography{io_amc}
\bibliographystyle{ieeetr}

\end{document}